\documentclass [11pt]{article}

\usepackage[dvips,pagebackref,colorlinks=true,citecolor=blue,linkcolor=blue,pdfpagemode=none,pdfstartview=FitH]{hyperref}

\usepackage{amssymb,amsmath}

\newcommand\R{{\mathbb R}}

\let\eps=\varepsilon
\def\epsilon{\varepsilon}
\def\phi{\varphi}

\newtheorem{theorem}{Theorem}
\newtheorem{lemma}[theorem]{Lemma}
\newtheorem{corollary}[theorem]{Corollary}

\newtheorem{claim}[theorem]{Claim}

\newenvironment{proof}{\noindent {\sc Proof:}}{$\Box$ \medskip}

\newcommand\E{\mathop{\mathbb E}\displaylimits}

\title {Max Cut and the Smallest Eigenvalue}

\author {{\large\sc Luca Trevisan}\thanks{{\tt luca@cs.berkeley.edu}.
U.C. Berkeley, Computer Science Division. This material is based upon 
work supported by the National Science Foundation under grant No.  CCF-0729137 
and by the BSF under grant 2002246.}}

\begin {document}

\sloppy

\maketitle

\begin{abstract}
We describe a new approximation algorithm for Max Cut.
Our algorithm runs in $\tilde O(n^2)$ time, where $n$ is the number of vertices, and
achieves an approximation ratio of $.531$.
On instances in which an optimal solution cuts a $1-\epsilon$ fraction of edges,
our algorithm finds a solution that cuts a $1-4\sqrt{\epsilon} + 8\epsilon-o(1)$ fraction of edges.

Our main result is a variant of spectral partitioning, which can be implemented in
nearly linear time. Given a graph in which the Max Cut optimum is a $1-\epsilon$
fraction of edges, our spectral partitioning algorithm
finds a set $S$ of vertices and a bipartition $L,R=S-L$ of $S$
such that at least a $1-O(\sqrt \epsilon)$ fraction of the edges incident on $S$ have
one endpoint in $L$ and one endpoint in $R$. (This can be seen as an analog of Cheeger's
inequality for the smallest eigenvalue of the adjacency matrix of a graph.)
Iterating this procedure yields the
approximation results stated above.

A different,  more complicated, variant of spectral partitioning leads to
an $\tilde O(n^3)$ time  algorithm that cuts $1/2 + e^{-\Omega(1/\eps)}$ fraction of edges in graphs
in which the optimum is $1/2 + \epsilon$.
\end {abstract}

\section{Introduction}

In the Max CUT problem, we are given an undirected graph with non-negative
weights on the edges and we wish to find a partition of the vertices (a {\em cut})
which maximizes the weight of edges whose endpoints are on different sides of
the partition (such edges are said to be {\em cut} by the partition). We
refer to the {\em cost} of a solution as the fraction of weighted edges of the graph
that are cut by the solution.

It is easy, given any graph, to find a solution that cuts half of the edges, providing
an approximation factor of $1/2$ for the problem.
The algorithm of Goemans and Williamson~\cite{GW94}, based on a 
Semidefinite Programming (SDP) relaxation, has a performance ratio
of $.878\cdots$ on general graphs, and it finds a cut of cost $1-O(\sqrt \epsilon)$
in graphs in which the optimum is $1-\epsilon$. 
Assuming the unique games conjecture, both results are best possible
for polynomial time algorithms \cite{K02:unique,KKMO04,MOO05} (see also \cite{ODW08}).
Arora and Kale~\cite{AK07} show that
the Goemans-Williamson SDP relaxation can be near-optimally solved in nearly linear time
in graphs of bounded degree (or more generally, in weighted graphs with bounded ratio
between largest and smallest degree). We show in Appendix \ref{sec:AK} that, using
a reduction \cite{T01}, the Arora-Kale algorithm can be used to achieve the approximation
performance of the Goemans-Williamson algorithm on all graphs in nearly-linear time.

A different rounding algorithm for the Goemans-Williamson relaxation, due
to Charikar and Wirth \cite{CW04}, finds a solution
that cuts at least a $1/2 + \Omega(\epsilon / \log 1/\epsilon)$ fraction of edges in graphs in which
the optimum is $1/2 + \epsilon$. This result too is tight, assuming the
unique games conjecture \cite{KOD06}.

No method other than SDP is known to yield an approximation better than $1/2$ for Max Cut,
and such approximation has been ruled out for large classes of Linear Programming
Relaxations \cite{VK07,SST07}.

A main source of difficulty in designing approximation algorithms for max cut is the
lack of good {\em upper bound} techniques for the max cut optimum of general graph.
Indeed,  suppose that one is able to design and analyse a new polynomial-time
algorithm for max cut achieving, say, a $.51$ approximation ratio, and consider
the behaviour of the algorithm when given a graph whose max cut optimum is $.501$.
Then the algorithm will clearly output a cut of cost $\leq .501$, but then
the computations performed by the algorithm, plus the proof of its approximation ratio,
provide a {\em certificate} that the optimum cut in the given graph is $\leq .501/.51 < .983$.
The problem is that, except for semidefinite programming, we know of no technique
that can provide, for every graph of max cut optimum $\leq .501$, a certificate
that its optimum is $\leq .99$. Indeed, the results of  \cite{VK07,SST07} show
that large classes of Linear Programming relaxations of max cut are unable to
distinguish such instances.

It is possible, however, to develop a new approximation algorithm that uses semidefinite
programming only in the analysis, by showing that if the algorithm outputs 
a cut of cost $c$, then there is a dual solution for the Goemans-Williamson
SDP relaxation of cost at most $c/.51$, thus proving that the max cut optimum
is at most $c/.51$ and that the algorithm has a performance ratio at least $.51$.
Such primal-dual algorithms, which use a relaxation only in the analysis, have been
derived for several problems based on {\em Linear Programming} relaxations, but
unfortunately, as discussed above, linear programming relaxations are unlikely to
be helpful in max cut approximation. As far as we know, the only examples of
primal-dual approximation algorithms for combinatorial problems based on
Semidefinite Programming are the algorithms for the sparsest cut
problem described in \cite{ARV04,KRV06,OSVV08}.

\subsection*{Our Results}

Our main result is a variant of the spectral partitioning algorithm with the
following property: given a graph $G=(V,E)$ in which the Max CUT optimum cost is $1-\epsilon$,
it finds a set $S$ and a partition of $S$ into
two disjoint sets of vertices $L,R$ such that the number of edges
with one endpoint in $L$ and one endpoint in $R$ is at least a $1-O(\sqrt{\epsilon})$
of the total number of edges incident\footnote{An edges $(i,j)$ is {\em incident} on
a set $S$ of vertices if at least one of the endpoints $i,j$ belongs to $S$.} on $S$.
More precisely, we show that the number of edges having both endpoints in $L$ or both
endpoints in $R$, plus half the number of edges having an endpoint in $S$ and
an endpoint in $V-S$ is at most a $2\sqrt \epsilon+o(1)$ fraction of the edges incident on $S$.
(See Theorem \ref{th:main} and the subsequent discussion.) We will ignore the $o(1)$ additive factors in the subsequent 
discussion in this section.

To derive an approximation algorithm for Max CUT, given a graph
we apply the partitioning algorithm and find sets $L,R$ as above, 
we remove the vertices in $L\cup R$
from the graph, recursively find a partition of the residual graph, and then
put back the vertices of $L$ on one side of the partition and vertices of $R$ on
the other side. This means that we cut all the edges that are cut in the recursive step,
plus all the edges with one endpoint in $L$ and one endpoint in $R$, plus at least
half of the edges between $S$ and $V-S$.
The recursion is stopped when less than half of the edges
incident on $S$ are cut, in which in case we return a greedy partition
of the residual graph.

We present an analysis of the recursive procedure due to Moses Charikar, which improves
an analysis of  ours which  appeared in a previous version of this paper.
The following observation plays an important role in the analysis:  at a generic step of the execution of
the algorithm, if the optimal solution in the original graph is $1-\epsilon$, and the current
residual graph holds a $\rho$ fraction of the original edges, then we know
that the optimum in the current residual graph is at least $1-\epsilon/\rho$,
and the spectral algorithm cuts at least a $1-2\sqrt{\epsilon/\rho}$ fraction
of the edges incident on $L\cup R$. When the recursion ends, it is because
the spectral algorithm cuts less than half of the edges incident on $L\cup R$,
and so the optimum of the residual graph at the end of the recursion
must be less than $15/16$, meaning that the
residual graph at the end of the recursion contains at most a $16\epsilon$ fraction
of the edges of the original graph. Putting together this information, a calculation shows
that the algorithm cuts at least a $1-4\sqrt \epsilon + 8\epsilon$ fraction
of edges of the graph. The ratio $(1-4\sqrt \epsilon + 8\epsilon)/(1-\epsilon)$
is always at least $.531$.

When applied to graphs in which the optimum is close to $1/2$ (in fact, to any graph
in which the optimum is smaller than $15/16$), our algorithm may simply return a greedy
partition. Thus, it fails to provide any non-trivial approximation to the Max CutGain problem,
which is the same as the Max Cut problem, except that we count the number of cut
edges {\em minus $|E|/2$}. (Equivalently, we count the number of cut edges minus
the number of uncut edges.) For Max CutGain we develop a more sophisticated spectral
partitioning algorithm with the following property: given a graph in which the
Max Cut optimum is $1/2+ \epsilon$, our algorithm finds sets $L,R$ such that
the number of edges incident on $L\cup R$ cut by the partition exceeds the number
of uncut edges by at least a $1/exp(\Omega(1/\epsilon))$ fraction of the edges incident on $L\cup R$.
Iterating this algorithm allows us to find a cut for the entire graph of
cost at least $1/2 + 1/exp(\Omega(1/\epsilon))$. 

This second algorithm can be also applied to the case in which edges have negative weights, and it
approximates a general class of quadratic programs. 
Given a symmetric real-valued matrix $Q$ with zeroes on the diagonal, if there exists
a vector $x\in \{ -1,1\}^V$ such that $x^T Qx \geq \epsilon \cdot ||Q||_1$, our algorithm finds
a vector $y\in \{-1,1\}^V$ such that $y^T Qy \geq exp(-O(1/\epsilon))\cdot ||Q||_1$, where
$||Q||_1 := \sum_{i,j} |Q(i,j)|$.
(The algorithm of Charikar and Wirth finds a vector $y$ such that
$y^T Qy \geq ||Q||_1 \cdot \epsilon / \log 1/\epsilon$.)

\subsection*{Relation to Cheeger's Inequality}

In the case of regular graphs, 
our main result, Theorem \ref{th:main}, may be seen as an analog of Cheeger's
inequality \cite{A86} for the smallest (rather than second largest) eigenvalue of the
adjacency matrix of the graph. We discuss this analogy in Section \ref{sec:cheeger}

\subsection*{Relation to the Goemans-Williamson Relaxation}

Our algorithm may also be seen as a primal-dual algorithm that produces, along
with a cut, a feasible solution to the semidefinite dual of the Goemans-Williamson
relaxation such that the cost of the cut is at least $.531$ times the cost of
the dual solution. We describe this view in Section \ref{sec:dual}.

\subsection*{Other Relations to Previous Work}

It has been known  that one can use spectral methods to 
certify an upper bound to the Max CUT optimum of a given graph. In
particular, if $G$ is a $d$-regular graph of adjacency matrix $A$,
and $M:= \frac 1d A$ has
eigenvalues $1=\lambda_1\geq \lambda_2 \geq \cdots \geq \lambda_n$, then one
can easily show\footnote{Inequality (\ref{eq:lambdaneasy}) appears to be a folklore
result. Lov\'asz \cite[Proposition 4.4]{L03} credits it to Delorme and Poljack \cite{DP93:ejc,DP93:mp}.
The earliest related reference we are aware of is  \cite[Theorem 2.1.4.i]{H79},
which states that if $V_1,V_2$ is a partition of a $d$-regular graph $G=(V,E)$, and
if $d_1$ is the average degree of the subgraph induced by $V_1$, then $n_1 d - nd_1 \leq -\lambda_n \cdot (n-n_1)$,
from which one can derive that $n_1 \cdot (d-d_1)$, the number of edges crossing the cut,
obeys $n_1 \cdot (d-d_1) \leq n_1 \cdot (n-n_1) \cdot (d-\lambda_n)/n$, and the latter
term is at most $n\cdot (d-\lambda_n)/4$.} that

\begin{equation}\label{eq:lambdaneasy} {\rm Max\ Cut} \leq \frac 12 + \frac 12 {|\lambda_n|}
\end{equation}

(Our Lemma \ref{lm:dual} is essentially a restatement of this fact.)

What is new is that we are able to prove a {\em converse}, in Lemma \ref{lm:main}, and
show that a non-trivial consequence follows whenever $|\lambda_n|$ is close to $1$.

As mentioned above, it was known that  $\lambda_n = -1$
if and only if $G$ has a bipartite connected component. In particular, if $G$ is
connected and not bipartite then $\lambda_n > -1$. Alon and Sudakov \cite{AS00}
consider the question of how small, in such case, can the gap $1-|\lambda_n|$ be.
They show that, if $G$ is connected and not bipartite, it has maximum degree
$d$ and  diameter $D$, and $\lambda_n$ is the smallest eigenvalue of the
adjacency matrix $A$, then $d-|\lambda_n| \geq  \frac{1}{(D+1)\cdot n}$. The bound
was improved to $d-|\lambda_n| \geq  \frac{1}{D\cdot n}$ by Cioaba \cite{C07}.
Our result implies the weaker bound $d-|\lambda_n| \geq \frac {1}{dn^2}$
in a $d$-regular graph.

The ``converse expander mixing lemma'' of Bilu and Linial \cite{BL06} has some
similarity with our approach to Max CutGain. Bilu and Linial show that
if $G$ is a $d$-regular graph, $A$ is the adjacency matrix, and
$\lambda_1\geq \cdots \geq \lambda_n$ are the eigenvalues of $M:=\frac 1d A$,
then if $\max \{ \lambda_2 , | \lambda_n | \} \geq \epsilon$ it follows
that there are sets $L,R$ such that the number of edges between $L$ and $R$
differs from what one would expect in a random $d$-regular graph by
a multiplicative error factor $\Omega(\epsilon / \log 1/\epsilon)$.
In our main result for Max CutGain (Theorem \ref{th:gain}) we have
a stronger assumption, that $|\lambda_n|\geq \epsilon$, but we need
to derive a much  stronger conclusion, namely that the number of edges
between $L$ and $R$ not only exceeds the number of edges that
one would expect in a random $d$-regular graph (a fact that can be
probably proved with the same quantitative result of Bilu-Linial), but
in fact exceeds the number of edges which are entirely contained in $L$
or entirely contained in $R$.

The main difference between our proof and the proof of Bilu and Linial is
that the combinatorial quantity that they relate to $\max \{ \lambda_2 , | \lambda_n | \} $
is the optimum of the normalized multilinear form $\max_{x,y \in \{-1,0,1\} } |x^T My|/(||x||\cdot||y||)$, for
a certain matrix $M$, while the combinatorial quantity that we wish to relate
to $|\lambda_n|$ is the optimum of the normalized 
homogeneous quadratic form $\max_{x \in \{-1,0,1\} } |x^T Mx|/||x||^2$, for a different
matrix $M$. Generally, it is considerably harder to round continuous relaxations
of quadratic forms of the latter type compared to multilinear forms of the first
kind. (See e.g. the introduction of \cite{CW04} and their discussion of their results
versus the results of Alon and Naor \cite{AN06}.)

The idea of iteratively removing parts of an instance in which one has a good
solution appears in various works on the sparsest cut problem (for example
in the way Spielman and Teng \cite{ST04} find a balanced separator using
their ``nibble'' procedure), and it was used to approximate the Max CUT problem (in
the version in which one wants to minimize the number of uncut vertices)
by Agarwal et al. \cite{ACMM05}. In the algorithm of Agarwal et al., as in 
our algorithm, the basic procedure that is being iterated finds a set $S$
of vertices and a bipartition $L,R$ of $S$ such that most of the edges
incident on $S$ have one endpoint in $L$ and one endpoint in $R$.

\section{Sparsification}
\label{sec:sparsify}

It follows from the Chernoff Bound that if we are given a graph $G=(V,E)$
and we sample $O(\delta^{-2} |V|)$ edges with replacement\footnote{If
the graph is unweighted, we sample from the uniform distribution over the edges;
otherwise we sample from the distribution in which each edge has a probability
proportional to its weight.} then, with high probability, every cut $(S,\bar S)$
has the same cost in the original graph as in the new graph, up to an additive
error $\delta$.\footnote{Note that the sparsified graph is an unweighted multigraph, and
that the sparsification process is considerably simpler than the one used
for algorithms for sparsest cut and other graph minimization problems.}

For this reason, all the dependency on $|E|$ in the running time of our algorithm
can be changed to a dependency on $|V|$ with an arbitrarily small loss in the
approximation factor.

\section{The Spectral Algorithm}

In this section we prove our main result.

\begin{theorem}[Main]\label{th:main} There is an algorithm that, given a graph $G=(V,E)$
for which the optimum of the Max CUT problem is at least $1-\epsilon$,  and a parameter $\delta$,
finds a vector $y \in \{-1,0,1\}^V$ such that

\[ \frac {\sum_{i,j}  A_{i,j}| y_i + y_j | } {\sum_i d_i |y_i|} \leq 4\sqrt{\epsilon} + \delta \]
where $A_{i,j}$ is the weight of edge $(i,j)$ and $d_i$ is the (weighted) degree of vertex $i$.

The algorithm can be implemented
in nearly-linear randomized time $O ( \delta^{-2} \cdot (|V|+|E|) \cdot \log |V|)$.
\end{theorem} 

To understand the statement of Theorem \ref{th:main}, let $y$ be the vector
returned by the algorithm, and call $L$ the set of vertices with negative coordinates in $y$,
and $R$ the set of vertices with positive coordinates. Then, up to constant factors,
the numerator counts the number of edges 
incident on $L\cup R$ which fail to have one
endpoint in $L$ and one endpoint in $R$, the denominator counts the number of
incident incident on $S$.  More specifically, the numerator counts four times
the edges that are entirely contained in $L$ or entirely contained in $R$, and twice
the edges that have one endpoint in $S$ and one endpoint in $V-S$. The denominator counts every edge
incident on $L\cup R$ once or twice, depending on whether one or both the endpoints of the
edge are in $S$.

The following form of the conclusion of Theorem \ref{th:main} will be convenient in our analysis: given
the vector $y$, call $M$ the number of edges incident on $L\cup R$, $U$ the number of ``uncut'' edges
that have both endpoints in $L$ or both endpoints in $R$, and $X$ the number of ``cross'' edges
that have exactly one endpoint in $L\cup R$; then

\[ U + \frac 12 X \leq \left( 2 \sqrt{\epsilon} + \frac \delta 2 \right) \cdot M\]

Let $A$ be the adjacency matrix of our input graph $G$ (hence $A_{i,j}$ is the weight
of the edge between $i$ and $j$), and $D$ be the diagonal matrix
such that $D_{i,i}$ is the weighted degree $d_i$ of vertex $i$ and $D_{i,j}=0$ for $i\neq j$.

Theorem \ref{th:main} follows by combining the following two results, and noting
that, for $a,b\geq 0$, $\sqrt{a+b} \leq \sqrt{a} + \sqrt{b}$.

\begin{lemma}\label{lm:dual}
If the optimum Max CUT in $G$ has cost at least $1-\epsilon$,
there is a vector $x\in \R^V$ such that

\[  x^T (D+A) x \leq 2\epsilon \cdot x^T D x \ . \]

Furthermore,  for every $\delta>0$, we can find in time $O (\delta^{-1}\cdot (|E| + |V|)\cdot \log |V|)$ a 
vector $x\in \R^V$ such that
\[  x^T (D+A) x \leq (2\epsilon +\delta) \cdot x^T D x \]
\end{lemma}

\begin{lemma} \label{lm:main}
Given a vector $x\in \R^V$ such that $x^T (D+A) x \leq \epsilon \cdot x^TDx$,  
we can find in time $O (|E| + |V|\log |V|)$ a vector $y\in \{-1,0,1\}^V$ such that

\begin{equation} \label{eq:lmmain} 
\frac {\sum_{i,j} A_{i,j} | y_i + y_j | } {\sum_i d_i |y_i|} \leq \sqrt{8\epsilon} \end{equation}
\end{lemma}

Lemma \ref{lm:dual} has a simple proof, and it can be seen as a statement
about the semidefinite dual of the Goemans-Williamson relaxation, as
discussed in Section \ref{sec:dual}. Lemma \ref{lm:main} is the main result of this
paper.

\subsection{Proof of Lemma~\protect\ref{lm:dual}}

Consider the optimization problem

\begin{equation} \label{eq:eig} \min_{x\in \R^V} \frac{x^TAx}{x^TDx} \end{equation}

Let $(S,\bar S)$ be an optimum cut for $G$, and define the vector $x^*\in \{ -1,1\}^V$
such that $x^*_i = 1$ if $i\in S$ and $x^*_i=-1$ otherwise. Then $x^{*T}Ax^*$ equals twice the difference
between the number of edges not cut by $(S,\bar S)$  and the number of edges that are
cut, which is at most $2\cdot (2\epsilon-1) \cdot |E|$. As for $x^{*T}Dx^*$, we have
\[ x^{*T}Dx^* = \sum_i d_i \cdot (x^*_i)^2 = \sum_i d_i = 2\cdot |E|\]
Thus $x^*$ is a feasible solution to (\ref{eq:eig}) of cost at most $2\epsilon-1$,
and if $\hat x$ is the optimal solution to (\ref{eq:eig}), then we must have

\[ \hat x^T A \hat x \leq (2\epsilon -1) \hat x^T D\hat x \]

To prove the ``furthermore'' part of the lemma, we observe that
the optimization problem in (\ref{eq:eig}) is equivalent to 

\begin{equation} \label{eq:eig:two} \min_{x\in \R^V} \frac{x^TD^{-1/2} AD^{-1/2} x}{x^Tx} \end{equation}

where $D^{-1/2}$ is the matrix that such that $D^{-1/2}_{i,j} = 0$ if $D_{i,j}=0$,
and $D^{-1/2}_{i,j} = 1/\sqrt{D_{i,j}}$ otherwise. In turn, the optimization
problem in (\ref{eq:eig:two}) is the problem of computing the smallest
eigenvalue of $D^{-1/2} A D^{-1/2}$, which is the same as computing
the largest eigenvalue of the positive semidefinite matrix $I- D^{-1/2}AD^{-1/2}$.

Given a $n\times n$ positive semidefinite matrix $M$ with $T$ non-zero entries
and of largest eigenvalue $\lambda_1$,
 and a parameter $\delta$,
it is possible to find a vector $x$ such that $x^TMx \geq \lambda_1\cdot (1-\delta) \cdot x^Tx$
in randomized time $O(\delta^{-1} \cdot (T+n) \cdot \log n)$ \cite{KW92}. Applying the 
algorithm to $I-  D^{-1/2} A D^{-1/2}$, which, as proved above, has
a largest eigenvalue which is at least $2-2\epsilon$, and which has $|E|+|V|$ non-zero
entries, we find in randomized time $O(\delta^{-1/2} \cdot (|E|+|V|) \cdot \log |V|)$
a vector $x'$ such that

\[ \frac {x'^T (I- D^{-1/2} AD^{-1/2}) x'}{x'^Tx'} \geq 2-2\epsilon - \delta \]
and so 
\[ x'^T D^{-1/2} AD^{-1/2} x' \leq (2\epsilon + \delta - 1) \cdot x'^Tx' \]

and, if we define $x'' := x'D^{1/2}$, then

\[ x''^T A x'' \leq (2\epsilon + \delta - 1) x''^TDx'' \]
which we can rewrite

\[ x''^T(A+D) x'' \leq (2\epsilon + \delta) x''^TDx'' \]

\subsection{Proof of Lemma~\protect\ref{lm:main}}

We now come to our main result.

The condition $x(D+A)x \leq \epsilon\cdot  xDx$ is equivalent to

\begin{equation} \label{eq:one}
 \frac 12 \sum_{i,j} A_{i,j} (x_i + x_j)^2 \leq \epsilon \sum_i d_i x_i^2 \end{equation}

Before starting the formal proof, we describe a heuristic
argument that gives some intuition for the actual proof.

\begin{description}
\item [\it Proof Idea.]  Equation (\ref{eq:one}) states
that the average value of $(x_i + x_j)^2$, for an edge $(i,j)$, is at most
$\epsilon$ times the average value of $x^2_i$ and $x^2_j$. So, non-rigorously,
we would guess that for a typical edges the value of $|x_i + x_j|$  is at
most about $\sqrt{\epsilon}$ times $|x_i| + |x_j|$. For this to happen, it must
be the case that $x_i$ and $x_j$ have different signs, and their absolute value
is nearly the same; that is, for some positive $c$, $x_i=-c$ and $x_j = c(1-\sqrt {\epsilon})$.
Suppose now that we pick a random threshold $t$, and we define $y_i=-1 \Leftrightarrow x_i \leq -t$
and $y_i=1 \Leftrightarrow x_i \geq t$. Then $|y_i-y_j|$ is 2 with probability $c\sqrt{\epsilon}$
and zero otherwise, while $|y_i|$ and $|y_j|$ are 1 with probability roughly $c$ and zero otherwise;
then it follows that the expectation of $\sum_{(i,j)} |y_i + y_j|$ is about a $\sqrt \epsilon$
fraction of the expectation of $\sum_i d_i |y_i|$.
\end{description}

Our algorithm, which we call the 2-Thresholds Spectral Cut algorithm and abbreviate 2TSC, is as follows:

\begin{itemize}
\item Algorithm 2TSC
\item For every vertex $k$
\begin{itemize}
\item Define the vector $y^k \in \{-1,0,1\}^V$ as follows:
 \[ \begin{array}{lcl} y^k_i  =  -1 & \mbox{iff} & x_i < -|x_k|\\
                   y^k_i =1     & \mbox{iff} & x_i > |x_k|\\
                   y^k_i = 0    & \mbox{iff} & |x_i| \leq |x_k|
  \end{array} \] 
\end{itemize}
\item Output the vector $y^k$ for which the ratio

\[ \frac{\sum_{i,j} A_{i,j} |y^k_i + y^k_j| } {\sum_i d_i |y^k_i|} \]
 is smallest
\end{itemize}

The algorithm can be implemented to run in $O(|E|+|V|\log |V|)$ time.
We first sort the vertices according to the value of $|x_i|$, and so we assume
we have $|x_1| \leq |x_2| \leq \cdots \leq |x_n|$ when we run 2TSC. At
each step $k$, we need to modify the vector $y$ only in positions $k$ and $k-1$,
and the cost of recomputing the ration is only  $O(d_{k} + d_{k-1})$, so that
all the $n$ steps together take time $O(|E|)$.

We need to argue that, under the assumption
of the Lemma, the algorithm outputs a vector $y$ such that
the ratio in (\ref{eq:lmmain}) is at most $\sqrt{8\epsilon}$

In order to analyze 2TSC, we study the following randomized process:

\begin{itemize}
\item Pick a value $t$ uniformly in $[0,\max_i x^2_i]$;
\item Define $Y \in \{-1,0,1\}^V$ as follows: 
\[ \begin{array}{lcl} 
 Y_i =-1 & \mbox{iff} & x_i < -\sqrt t\\
 Y_i =1 & \mbox{iff}  & x_i > \sqrt t\\ 
 Y_i = 0 & \mbox{iff} & |x_i| \leq \sqrt t
\end{array} \] 
\end{itemize}

Every $Y$ that is generated by the probabilistic process with positive probability is considered by
algorithm 2TSC at some stage; this implies that if algorithm 2TSC 
outputs a vector $y$ such that $\sum_{i,j} A_{i,j} |y(i)+y(j)| > \sqrt{8\epsilon} \sum_i d_i |y_i|$,
then in the randomized process we must
have $\sum_{i,j} A_{i,j} |Y(i)+Y(j)| > \sqrt{8\epsilon} \sum_i d_i |Y_i|$ with probability 1 and, in particular,
$\E \sum_{i,j} A_{i,j} |Y(i)+Y(j)|> \sqrt{8\epsilon} \E \sum_i d_i |Y_i|$. 

We shall prove that
$\E \sum_{i,j} A_{i,j} |Y(i)+Y(j)| \leq \sqrt{8\epsilon} \E \sum_i d_i |Y_i|$ and so we shall conclude that
the output of algorithm 2TSC satisfies the Claim.

Since Equation (\ref{eq:one}) and the distribution $Y$ are invariant under multiplying $x$
by  a scalar, we may assume that
$\max_i |x_i|= 1$, so that $t$ is chosen uniformly in $[0,1]$.

A case analysis shows that, for every edge $(i,j)$,

\begin{equation} \label{eq:two} \E |Y_i+Y_j| \leq |x_i + x_j | \cdot ( |x_i| + |x_j| ) 
\end{equation}

To verify Equation (\ref{eq:two}) we need to distinguish the case in which $x_i$ and $x_j$ have
different signs from the case in which they have the same sign. We assume without
loss of generality that $|x_i| > |x_j|$.

\begin{itemize}
\item If they have different signs, and, say, $|x_i| > |x_j|$, then $|Y_i + Y_j|=1$
when $|x_j|^2 \leq t \leq |x_i|^2$, and zero otherwise. Indeed, 
if $t< |x_j|^2$, then $Y_i=-Y_j$ and $|Y_i+Y_j|=0$,
and if $t> |x_i|^2$ then $Y_i=Y_j=0$.

So $\E |Y_i+Y_j|$ equals $|x_i|^2 - |x_j|^2$, which is equal to the right-hand
side of  Equation (\ref{eq:two}).

\item If they have the same sign, then $|Y_i+Y_j|=2$ when $t \leq |x_j|^2$,
$|Y_i + Y_j|=1$ when $|x_j|^2 < t \leq |x_i|^2$, and $|Y_i+Y_j|=0$ when $t> |x_i|^2$.

Overall, $\E |Y_i+Y_j|$  equals $2x_j^2 + (x_i^2-x_j^2) = x_j^2 + x_i^2$. The
right-hand-sise of Equation (\ref{eq:two}) is $(x_i + x_j)^2$, which is only larger.
\end{itemize}

Note also that $\E |Y_i| = x_i^2$.

To complete our argument it remains to apply Cauchy-Schwarz and standard manipulations.

\begin{eqnarray*}
 \E \sum_{i,j} A_{i,j} |Y_i + Y_j|  
& \leq & \sum_{i,j} A_{i,j} |x_i + x_j| \cdot (|x_i| + |x_j|) \\
& \leq &  \sqrt{\sum_{i,j} A_{i,j} |x_i + x_j|^2} \cdot \sqrt {\sum_{i,j} A_{i,j} (|x_i| + |x_j|)^2 }
\end{eqnarray*}

By our assumption,

\[ \sum_{i,j} A_{i,j} |x_i + x_j|^2 \leq 2 \epsilon \sum_{i} d_i x_i^2 \]

and it is a standard calculation that

\[ \sum_{i,j} A_{i,j} (|x_i| + |x_j|)^2 \leq 2\sum_{i,j} A_{i,j} (|x_i|^2 + |x_j|^2) = 4\sum_i d_i x_i^2 \]

and so

\[  \E \sum_{i,j} A_{i,j} |Y_i + Y_j| \leq \sqrt {8\epsilon} \sum_i d_i x_i^2 = \sqrt {8\epsilon} \E \sum_i d_i |Y_i| \]

This completes the proof that Algorithm 2TSC performs as required by the Lemma.

\section{Approximation for Max Cut}

In this section we analyze the following algorithm

\begin{itemize}
\item Algorithm: {\sc Recursive-Spectral-Cut}
\item Input: graph $G=(V,E)$, accuracy parameter $\delta$
\item Run the algorithm of Theorem \ref{th:main} with accuracy parameter $\delta$, 
and let  $y\in \{-1,0,1\}$ be the solution found by the algorithm; call 
$M$ the weighted number of edges $(i,j)$ such that least one of $y_i$ or $y_j$ is non-zero,
$C$ the weighted number of {\em cut} edges $(i,j)$ such that $y_i,y_j$ are both non-zero and
have opposite signs, and $X$ the weighted number of {\em cross} edges $(i,j)$ such that
exactly one of $y_i,y_j$ is zero;
\item If $C+\frac 12 X \leq \frac 12 M$, then find a partition of $V$ that cuts $\geq |E|/2$ edges, and return it.
\item If $C+\frac 12 X  > \frac 12 M$, then let $L:=\{ i: y_i=-1\}$, $R:=\{ i: y_i=1\}$, $V':= \{i:y_i=0\}$,
let $G'=(V',E')$ be the graph induced by $V'$,
recursively call {\sc Recursive-Spectral-Cut} on $G'$, and let $V_1,V_2$ be the 
partion found by the algorithm; return 
$(V_1 \cup L, V_2 \cup R)$ or $(V_1\cup R, V_2\cup L)$, whichever is better.
\end{itemize}

Note that the algorithm runs in randomized time $O ( \delta^{-2} \cdot |V| \cdot (|V|+|E|) \cdot \log |V|)$ because
each iteration takes time $O ( \delta^{-1} \cdot (|V|+|E|) \cdot \log |V|)$ and there are
at most $|V|$ iterations.

In a preliminary version of this paper
we presented a simple argument showing that if $opt\geq 1-\epsilon$, then
the algorithm cuts at least $1-O(\epsilon^{1/3})-\delta$ fraction of edges.
The following tighter argument is due to Moses Charikar (personal communication, July 2008).

\begin{theorem} \label{th:largecut}
If Algorithm {\sc Recursive-Spectral-Cut} receives in input a graph $G=(V,E)$ whose optimum is $1-\epsilon$,
with $\epsilon< 1/16$
then it finds a solution that cuts at least a $1-4\sqrt{\epsilon} + 8 \epsilon -\frac \delta 2$ fraction of edges.
\end{theorem}

\begin{proof} 
Consider the $t$-th iteration of the algorithm, and let $G_t$ be the residual graph
at that iteration, and let $\rho_t \cdot |E|$ be the number of edges of $G_t$. Then
we observe that the Max Cut optimum in $G_t$ is at least $1-\epsilon/\rho_t$.

Let $S_t$ be the set of vertices and $L_t,R_t$ the partition
found by the algorithm of Theorem \ref{th:main}. Let 
$G_{t+1}$ be the residual graph at the following step, and $\rho_{t+1}\cdot |E|$
the number of edges of $G_{t+1}$. (If the algorithm stops at the $t$-th iteration, we shall
take $G_{t+1}$ to be the empty graph; if the algorithm discards $L_t,R_t$ and chooses a
greedy cut, we shall take $G_{t+1}$ to be empty and $L_t,R_t$ to be the partition given
by the greedy cut.)

We know by Theorem \ref{th:main} that the algorithm
will cut at least a $1-2\sqrt{\epsilon/\rho_t} - \delta/2$ fraction of
the $|E|\cdot (\rho_t - \rho_{t+1})$ edges incident on $S_t$.

Indeed, we know that at least a $\max \{ 1/2 , 1-2\sqrt{\epsilon/\rho_t} -\delta/2 \}$ fraction of
those edges are cut (for small value of $\rho_t$, it is possible that 
$1-2\sqrt{\epsilon/\rho_t} + \delta/2 < 1/2$, but the algorithm is always guaranteed to cut
at least half of the edges incident on $S_t$). This means that any convex combination
of $1/2$ and $1-2\sqrt{\epsilon/\rho_t} - \delta/2$ is still a lower bound on the fraction
of edges incident on $S_t$ cut by the algorithm.

If both $\rho_t$ and $\rho_{t+1}$ are at least $16\epsilon$, we are going to use the lower bound

\begin{eqnarray*}
\displaystyle 
|E| \cdot (\rho_t - \rho_{t+1}) \cdot \left( 1 - 2\sqrt{ \frac \epsilon {\rho_t} } - \frac \delta 2\right)
& =&  |E| \int_{\rho_{t+1}}^{\rho_t} \left( 1 - 2\sqrt{ \frac \epsilon {\rho_t} } - \frac \delta 2 \right) d r\\
\displaystyle & \geq & 
|E| \int_{\rho_{t+1}}^{\rho_t} \left( 1 - 2\sqrt{ \frac \epsilon {r} } + \frac \delta 2\right) d r 
\end{eqnarray*}

If $\rho_t \geq 16\epsilon \geq \rho_{t+1}$, then we use the lower bound

\[ |E| \cdot (\rho_t - 16\epsilon ) \cdot \left( 1 - 2\sqrt{ \frac \epsilon {\rho_t} } + \frac \delta 2\right)
+ |E| \cdot (16 \epsilon - \rho_{t+1}) \cdot \frac 12 
\geq |E| \int_{16 \epsilon}^{\rho_t} \left( 1 - 2\sqrt{ \frac \epsilon {r} } - \frac \delta 2\right) d r 
+ |E| \cdot \int_{\rho_{t+1}}^{16 \epsilon}  \frac 12 dr \]

Finally, if both $\rho_t$ and $\rho_{t+1}$ are smaller than $16\epsilon$, we use the lower bound

\[ |E| \cdot (\rho_t - \rho_{t+1}) \cdot \frac 12 = |E| \cdot \int_{\rho_{t+1}}^{\rho_t} \frac 12 dr \]

Summing those bounds, we have that the number of edges cut by the algorithm is at least 

\[ |E| \cdot \left( \int_{16\epsilon}^1 \left( 1 - 2\sqrt{ \frac \epsilon {r} } - \frac \delta 2\right) d r
+ \int_0^{16\epsilon} \frac 12 dr \right) =  |E| \cdot \left( 1 - 4\sqrt{\epsilon} + 8 \epsilon - (1-16\epsilon) \frac \delta 2 \right)\]
\end{proof}

\begin{corollary}
Algorithm {\sc Recursive-Spectral-Cut} is a $.531128 - \delta$ approximate algorithm for Max Cut.
\end{corollary}

\begin{proof}
Write $opt=1-\epsilon$. If $\epsilon> 1/16$ then the algorithm finds a solution
of cost $>1/2$ and the approximation ratio is $16/30 > 5.33333$.

If $1/16 \leq \epsilon \leq 1/2$, then the algorithm finds a solution
of cost at least $1-4\sqrt{\epsilon} + 8 \epsilon-\delta/2$, and the approximation ratio is
at least

\[   \frac { 1- 4\sqrt{\epsilon} + 8 \epsilon-\delta/2}{1-\epsilon} \geq
\frac { 1- 4\sqrt{\epsilon} + 8 \epsilon}{1-\epsilon} - \delta\]

If we call $\rho(\epsilon) := \frac { 1- 4\sqrt{\epsilon} + 8 \epsilon}{1-\epsilon}$, then 
some calculus shows that, for $1/16 \leq \epsilon \leq 1/2$, $\rho(\epsilon)$
is minimized at $.05496$ (the smallest root of $-2x^2 +9x-2=0$)
and is always at least $.531128\cdots$.
\end{proof}

\section{Relation to Cheeger's Inequality}
\label{sec:cheeger}

In this section we compare our main result, Theorem \ref{th:main}, with
Cheeger's inequality \cite{A86}. We restrict our discussion to the case of regular graph.

If $G$ is a $d$-regular graph,  $A$ is its adjacency matrix, and $M:= \frac 1d A$, then 
$M$ has $n$ eigenvalues, counting multiplicities, which we shall
call $\lambda_1 \geq \lambda_2 \geq \cdots \geq \lambda_n$. 
It is always the case that $\lambda_1 = 1$, and that $|\lambda_i | \leq 1$ for every $i$.
The extremal cases are captured by the following well-known facts:

\begin{enumerate}
\item $\lambda_2 = 1$ if and only if $G$ is disconnected, that is, if and only if there 
is a set $S$, $|S| \leq |V|/2$, such that no edge of $G$ leaves $S$.
\item $\lambda_n = -1$ if and only if $G$ contains a bipartite connected component, that is,
if and only if there is a set $S$ and partition of $S$ into disjoint sets $L,R$, such that
all edges incident on $S$ have one endpoint in $L$ and one endpoint in $R$.
\end{enumerate}

Cheeger's inequality characterizes the cases in which $\lambda_2$ is close to $1$
as those in which there is a  set $S$, $|S| \leq |V|/2$ such that
the number of edges between $S$ and $V-S$ is small compared to $d|S|$.

If we define $h(G)$ to be the {\em edge expansion} of $G$,

\[ h(G) = \min_{S\subseteq V: \ |S| \leq |V|/2} \frac {edges(S,V-S)}{d|S|} \]

then we have Cheeger's inequality

\begin{equation} \sqrt{2 \cdot (1-\lambda_2) } \geq  h(g) \geq \frac 12 \cdot (1-\lambda_2 ) 
\end{equation}

Similarly, Lemmas \ref{lm:dual} and \ref{lm:main} characterizes the cases in which $\lambda_n$ is close to $-1$
as those in which there is a set $S$ and a partition $(L,R)$ of $S$ such that the number of edges incident 
on $S$ which fail to be cut by the partition is small compared to $d|S|$.

Define the {\em bipartiteness ratio} number of a graph to be 

\[ \beta(G) :=  \min_{y\in \{-1,0,1\}^V } \frac {\sum_{i,j} |y_i + y_j | }{2d\sum_i |y_i|}\]

which is equivalent to

\[ \beta(G) = \min_{S\subseteq V, \ (L,R)\ {\rm partition\ of}\ S} \frac{2edges(L) + 2edges(R) + edges(S,V-S)}{d|S|}
\]

then we have

\begin{equation} \label{eq:bipcheeger} \sqrt{2 \cdot (1-|\lambda_n|)} \geq \beta(G) 
\geq \frac 12  \cdot (1-|\lambda_n|)
\end{equation}

There are examples in which both inequalities in (\ref{eq:bipcheeger}) are tight within
constant factors.

If
we take an odd cycle with $n$ vertices, then $\beta(G) \geq \frac 1n$, because
for every subset $S$ of vertices and for every bipartition of $S$ there is at least
one failed edge, and the number of edges incident on $S$ is at most $n$. In an
odd cycle, however, $d=2$ and $|\lambda_n | = 2-O(1/n^2)$, and so $\beta$
is as large as $\Omega(\sqrt{1-|\lambda_n|})$.

To see the tightness of the other inequality, start from a $k$-regular expander
such that, say, $\max \{ \lambda_2 , |\lambda_n| \} \leq 1/2$. (Such graphs exist
for constant $k$.) Then construct $G$ by taking the disjoint union of the edges
of $G$ and the edges of a $k\cdot(1-\epsilon)/\epsilon$-regular bipartite graph,
so that the resulting graph is $d$-regular with $d:= k/\epsilon$. There is a cut that
cuts all the edges of the bipartite graph, so $\beta(G) \leq \epsilon$,
but the smallest eigenvalue of $M$ is at least $-1+k/2d \geq -1+\epsilon/2$,
meaning that $\beta$ is $O(1-|\lambda_n(G)|)$.

Our results, as stated in  (\ref{eq:bipcheeger}), are not just syntactically similar
to Cheeger's inequality:
There are also similarities between the proof of Cheeger's inequality
and of Theorem \ref{th:main}. The analysis in Cheeger's inequality
relies on the study of the quadratic form

\begin{equation} \label{cheegerform} \sum_{i,j} A(i,j) \cdot (x_i - x_j )^2 \end{equation}

and it is based on the intuition that if (\ref{cheegerform}) is small compared to $\sum_i x_i^2$ then
for most edges $(i,j)$ we have $x_i \approx x_j$.

Our analysis was based on the study of the quadratic form

\begin{equation} \label{betaform} \sum_{i,j} A(i,j) \cdot (x_i + x_j )^2 \end{equation}

and the intuition that if (\ref{betaform}) is small compared to $\sum_i x_i^2$ then
for most edges we have $x_i \approx - x_j$.

\section{Relation to the Goemans-Williamson Relaxation}
\label{sec:dual}

The dual of the Goemans-Williamson relaxation is

\begin{equation}
\begin{array}{l}
\min |E| - \frac 14 \sum_i y_i\\
\mbox{subject to}\\
D+A - diag(y_1,\ldots,y_n) \succeq 0
\end{array}
\label{gwdual}
\end{equation}

We can see  Lemma \ref{lm:dual} as 
stating a special case of the weak duality fact that the cost of every
feasible solution to (\ref{gwdual}) is an upper bound to
the optimal cut in the graph.

Indeed, if the optimal cut is of size $> |E|\cdot (1-\epsilon)$, then
no solution of cost $\leq |E| \cdot (1-\epsilon)$ can be feasible
for (\ref{gwdual}). In particular, the solution $y_i = 2\epsilon d_i$
has cost $1-\epsilon$ and cannot be feasible, meaning that
$D(1-2\epsilon) +A$ cannot be feasible, and there is a vector $x$
such that $x (D(1-2\epsilon) + A) x <0$.

In turn, Lemma \ref{lm:main} has the following primal dual interpretation:
given a graph $G$, there is an $\epsilon$ such that algorithm 2TSC finds
$L,R$ such that  $C+ \frac 12 X \geq (1-2\sqrt \epsilon -\delta/2) M$,
and the solution $y_i := 2\epsilon d_i$ is feasible
for (\ref{gwdual}), thus showing that the Max Cut optimum is at most $1-\epsilon$.

Given this premise, we can now view algorithm {\sc Recursive-Spectral-Cut} as a primal-dual
algorithm. 

At step $t$ of the recursion, let $\rho_t|E|$ be the number of edges in
the residual graph $G_t$, and $C_t$ and $X_t$ be the number of cut and cross edges
in the solution $L_t,R_t$ found by the algorithm. Define $\epsilon_t$
so that $1-\epsilon_t/\rho_t$ is the upper bound on the Max Cut of $G_t$
given by the dual solution associated to the algorithm as above, and
the algorithm satisfies $C_t + \frac 12 X_t \geq (1-2\sqrt{\epsilon_t/\rho_t}-\delta/2)M_t$.
Then the dual solution at time $t$ also proves an upper bound $1-\epsilon_t$
to the Max Cut optimum of $G$. Let $\epsilon := \max_t \epsilon_t$; then
we have (i) a dual solution proving that the Max Cut of $G$ is $\leq 1-\epsilon$,
and we know that (ii) at every step $t$ we have $C_t + \frac 12 X_t \geq (1- 2\sqrt{\epsilon/\rho_t}-\delta/2)M_t$.
From fact (ii) and the analysis done in the proof of Theorem \ref{th:main} we
see the algorithm outputs a solution that cuts at least
a $1-4\sqrt{\epsilon} + 8\epsilon - \delta/2$ fraction of edges, and it
is able to output a feasible dual solution to the GW relaxation
proving a $1-\epsilon$ upper bound to the optimum.

In particular, the ratio between the cost of the solution found by the algorithm
and the upper bound provided by the dual solution is always at least $.531$.

\section{Quadratic Programming and the  Max CutGain Problem}

Let $A$ be the adjacency matrix of a weighted graph with no self-loops, possibly with negative
weights, let
$d_i := \sum_j |A_{i,j}|$ be the weighted degree of node $i$, and $D:= diag(d_1,\ldots,d_n)$. 
{\em Max-Cut Gain} is the optimization problem

\begin{equation} \max_{ y\in \{ -1,1 \}^V } - \frac{y^T A y}{y^T D y} \end{equation}

In words, Max Cut Gain is the maximum, over all cuts, of the difference between the number
of cut edges and the number of edges that are not cut, divided by the total number
of edges. Equivalently, the optimum of Max Cut Gain is $\epsilon$ if and only if
the optimum of Max Cut is $\frac 12 + \frac 12 \epsilon$. (The name of the problem
comes from the fact that one is measuring how much one {\em gains} by using an optimum
cut compared to a random cut, which only cuts a $1/2$ fraction of edges.)

Note that, up to the scaling that we do by dividing by $y^TDy = \sum_i d_i$, we are
considering the problem

\begin{equation} \max_{ y\in \{ -1,1 \}^V } y^T Q y \end{equation}
where $Q$ is an arbitrary symmetric matrix with zeroes on the diagonal. Apart
from the restriction to symmetric matrices, this is the same family
of quadratic programs studied by Charikar and Wirth \cite{CW04}. It
helps intuition, however, to continue to think about $A=-Q$ as the adjacency
matrix of a weighted undirected graph.

We define the {\em gain ratio} of a graph the quantity

\begin{equation} \gamma(G):= \max_{ y\in \{ -1,0,1 \}^V } - \frac{y^T A y}{y^T D y} \end{equation}

In the {\em gain ratio}, we consider all subsets $S\subseteq V$ of vertices, and all
partitions $(L,R=S-L)$ of the set $S$; the objective function is the ratio between
twice the difference of cut edges minus uncut edges among the edges induced by $S$, divided
by the volume of $S$. If one imposed the additional constraint
 $S=V$, then one would recover the Max Cut Gain problem.

Let $\lambda_n$ be the smallest eigenvalue of the matrix $M:= D^{-1/2} A D^{-1/2}$; then
we see that

\begin{equation} \gamma(G) \leq |\lambda_n| \end{equation}

because

\[ |\lambda_n| = - \min_{z\in \R^V} \frac{z^T Mz}{z^Tz} = \max_{x\in \R^V} - \frac{x^T A x}{x^TDx} 
\geq  \max_{ y\in \{ -1,0,1 \}^V } - \frac{y^T A y}{y^T D y} = \gamma(G) \]

we conjecture that

\begin{equation} \gamma(G) \geq \Omega \left( \frac{|\lambda_n|} {\log  \frac 1 {|\lambda_n|}} \right) \end{equation}

but we are only able to prove the considerably weaker result that $\gamma(G) \geq e^{-O(1/|\lambda_n|)}$.

We use the following approach. Let $x\in \R^V$ be a real vector, and $Y$ be a distribution
over discrete vectors $\{ -1,0,1\}^V$. We say that $Y$ is a $(c_1,c_2,\delta)$-good (randomized) rounding of $x$ if

\begin{enumerate}
\item $| c_1 \cdot \E Y_i Y_j - x_ix_j | \leq \delta \cdot (x_i^2 + x_j^2)$
\item $\E |Y_i| \leq c_2 x_i^2$
\end{enumerate}

We have the following simple fact:

\begin{claim}
If  $x$ is a vector such that $-x^T Ax \geq \epsilon\cdot x^TDx$, and $Y$ is a 
a $(c_1,c_2,\delta)$-good rounding  of $x$, then the support of $Y$
contains a vector $y\in \{-1,0,1\}^V$ such that

\[ - y ^T A y \geq \frac 1 {c_1c_2}(\epsilon - 2\delta) \cdot y^T D y \]
\end{claim}

\begin{proof}
We have
\[ \E \sum_{i,j} - A_{ij} Y_iY_j \]
\[ \geq \frac 1{c_1} \left( \sum_{ij} -A_{i,j} x_ix_j \right) +\frac 1 {c_1} 2\delta \sum_i d_i x_i^2 \]
\[ \geq \frac 1{c_1} \cdot (\epsilon - 2\delta) \sum_i d_i x_i^2\]
\[ \geq \frac 1{c_1c_2} \cdot (\epsilon - 2\delta) \E \sum_i d_i |Y_i| \]

and so

\[ \frac{ \E  \sum_{i,j} - A_{ij} Y_iY_j }{\E \sum_i d_i |Y_i|} \geq \frac 1 {c_1c_2} (\epsilon-2\delta) \]
and in particular there must exist a vector $y\in \{-1,0,1\}$ such that

\[ \frac{ \sum_{i,j} - A_{ij} y_iy_j }{\sum_i d_i |y_i|} \geq \frac 1 {c_1c_2} (\epsilon-2\delta) \]
\end{proof}

\begin{lemma}[Main] \label{lm:gainmain}
For every $x\in \R^V$ and every $\ell > 1$ there is a $(c_1,c_2,1/\ell)$-good rounding
of $x$ such that $c_1\cdot c_2 \leq \ell^{-1} \cdot e^{\ell}$.
\end{lemma}

\begin{proof}
Given $x$, we assume without loss of generality that $|x_i|\leq 1$ for every $i$,
and we consider the following distribution $Y$:

\begin{itemize}
\item Pick a threshold $t\in [0,1]$ so that $t^2$ is uniformly distributed in $[0,1]$;
\item For every vertex $i$, pairwise independently:
\begin{itemize}
\item If $|x_i| > t$ or $|x_i| < t\cdot e^{-\ell}$, then set $Y_i := 0$;
\item If $t \cdot e^{-\ell} \leq |x_i| \leq t$, then set $Y_i:=sign(x_i)$ with probability $|x_i|/t$,
and $Y_i:=0$ with probability $1-|x_i|/t$.
\end{itemize}
\end{itemize}

We begin with the calculation of the expectations $\E |Y_i|$.

\begin{claim} \label{cl:ctwo}
$\E |Y_i| = 2\cdot (e^\ell - 1) \cdot x_i^2$
\end{claim}

\begin{proof} [Of Claim \label{cl:two}]
The threshold $t$ is chosen according to a distribution whose density function is $2t$
for $t\in [0,1]$; conditioned on a specific choice of $t$, the expectation of $|Y_i|$
is 0 if $|x_i| >t$ or $|x_i| < t e^{-\ell}$, and it is $|x_i|/t$ otherwise. Hence, we have

\[ \E |Y_i| = \int_{|x_i|}^{|x_i|e^{\ell}} 2t  \cdot \frac {|x_i|}{t} dt =
\int_{|x_i|}^{|x_i|e^{\ell}} 2 |x_i| dt = 2\cdot (e^\ell - 1) \cdot x_i^2  \]
\end{proof}

Claim \ref{cl:two} tells us that we can take $c_2 = 2 \cdot (e^\ell - 1) \leq 2e^\ell$. 
The following two claims  give us that we can take $c_1 = 1/2\ell$, so that
$c_1c_2 \leq \frac 1 \ell \cdot e^\ell$ as required.

\begin{claim} \label{cl:conea}
If $|x_i| > e^{\ell} |x_j|$, then, for every $c$

\[ | c \E Y_iY_j - x_ix_j | \leq \frac 1{\ell} x_i^2 \] 
\end{claim}

\begin{proof}[Of Claim \ref{cl:conea}] 
Just note that, under the assumption of the claim, $\E Y_iY_j = 0$,
and $|x_ix_j| \leq e^{-\ell} x_i^2 \leq \ell^{-1} x_i^2$.
\end{proof}

\begin{claim} \label{cl:coneb}
If $|x_j| \leq |x_i| \leq  e^{\ell} |x_j|$, then
\[ \left| \frac 1 {2\ell} \cdot \E Y_iY_j - x_i x_j \right| \leq \frac 1 \ell  \cdot x_i ^2\]
\end{claim}

\begin{proof}[Of Claim \ref{cl:coneb}] 
Consider the expectation of $Y_iY_j$, $i\neq j$, conditioned on a fixed choice of $t$.

$Y_iY_j=0$ whenever $|x_i|\geq t$ or $|x_j| \leq te^{-\ell}$.
If $t$ is such that $|x_i| \leq t \leq |x_j| e^{\ell}$, then the conditional expectation
of $Y_iY_j$ is $x_i x_j /t^2$. Overall, we have
\[ \E Y_jY_j = \int_{|x_i|}^{|x_j| e^{\ell} } 2t \cdot \frac{ x_i x_j }{t^2} dt
= 2 x_i x_j \cdot \int_{|x_i|}^{|x_j| e^{\ell} } \frac 1t dt = 
2 x_i x_j\cdot \left( \ell  - \ln \frac{|x_i|}{|x_j|}
 \right)  \]
So we have
 \begin{eqnarray*} 
\displaystyle \left| \frac 1 {2\ell} \cdot \E Y_iY_j - x_i x_j \right| 
& = & |x_ix_j| \cdot \frac 1 \ell \cdot \ln  \frac{|x_i|}{|x_j|}\\
\displaystyle & = & x_i^2 \cdot \frac 1 \ell \cdot \frac{|x_j|}{|x_i|} \cdot  \ln  \frac{|x_i|}{|x_j|}\\
\displaystyle  & \leq & x_i^2 \cdot \frac 1 \ell
\end{eqnarray*} 
where the last inequality follows from the fact that $\rho \ln \frac 1 {\rho} \leq 1$ for
every $0 < \rho \leq 1$.
\end{proof}

The lemma now follows.
\end{proof}

In order to make the proof constructive, we need to show that we can find
a vector $y$ in the sample space of $Y$ as in the conclusion of the lemma.
Suppose that the distribution of $Y$ described above is such that
$- \E Y^T A Y \geq \E \delta Y^T D Y$. 

A first observation is that there must be a threshold $t^*$ such that,
conditioned on that particular choice of $t$, we still have
$- \E [Y^T A Y | t=t^*] \geq \delta \E [Y^T D Y | t=t^*]$. Once we find such a threshold,
we can search in the sample space of $Y|t=t^*$, which is of polynomial size.

It remains to describe how to find a threshold $t^*$ as above.
Let us say that two thresholds $t_1,t_2$ are {\em combinatorially indistinguishable}
if the sets of vertices $\{ i: \delta t_1 \leq |x_i| \leq t_1 \}$ and $\{ i: \delta t_2 \leq |x_i| \leq t_2 \}$
are equal, and call $S$ the set of vertices.

Then we have 

\[ - \frac{\E [Y^T A T | t=t_1]}{\E [ Y^TDY | t=t_1]} = 
- \frac{\sum_{i,j \in S} A_{ij} x_i x_i/t_1^2}{\sum_{i\in S} d_i |x_i|/t_1} 
= - \frac 1 {t_1} \cdot \frac{\sum_{i,j \in S} A_{ij} x_i x_i}{\sum_{i \in S} d_i |x_i|} \]
and, similarly
\[ - \frac{\E [Y^T A T | t=t_1]}{\E [ Y^TDY | t=t_2]} 
= - \frac 1 {t_2}\cdot \frac {\sum_{i,j \in S} A_{ij} x_i x_i}{\sum_{i \in S} d_i |x_i|} \]

so that it is always preferable to choose the smaller threshold. This means that for every equivalence
class of combinatorially indistinguishable thresholds we only need to look at one of them, in order to 
find $t^*$, and so we only need to consider at most $2|V|$ thresholds. In particular, $t^*$ can be
found in $O(|E|+|V|)$ time. A nearly pairwise independent sample space of size $\tilde O(|V|)$ can
be used instead of a perfectly pairwise independent one so that the whole algorithm takes time
$\tilde O(|V|+|E|)$, at the price of a $o(1)$ additive loss in the approximation.

The following theorem summarizes our progress so far.

\begin{theorem}\label{th:gain}
There is a nearly quadratic time algorithm that in input a graph $G=(V,E)$ such that $\gamma(G)\geq \epsilon$
finds a set $S$ and a partition $(L,R)$ of $S$ whose gain is at least $e^{-\Omega(1/\epsilon)}$. 
\end{theorem}

\begin{proof} We call the algorithm {\em Four-Threshold Spectral Cut}, or 4TSC.

\begin{itemize}

  \item Algorithm 4TSC

  \item Input: Graph $G=(V,E)$
  \begin{itemize}

    \item Let $A$ be the adjacency matrix of $G$, $D$ be the matrix of degrees, $M:= D^{-1/2} M D^{-1/2}$. Find 
a vector $x\in \R^V$ such that $\epsilon := -x^TMx/x^Tx \leq 2|\lambda_n|$, where $\lambda_n$ is the smallest eigenvalue
of $M$. Set $\ell = 10/\epsilon$

    \item For every threshold $t$ in the set $\{ x(i) : i\in V \} \cup \{ e^{-\ell} x(i) : i \in V \}$

    \begin{itemize}
      \item Let $Y_1,\ldots,Y_n$ be a distribution of sample space $\Omega_t$ that is $\epsilon/10$-close to
           pairwise independence, and such that
            $Y_i\equiv 0$ if $|x_i|>t$ or $|x_j| <e^{-\ell} t$; and such that
            $Y_i=sign(x_i)$ with probability $|x_i|/t$ otherwise.
    \end{itemize}
    \item Output the vector $y$ in the union of $\Omega_t$ that maximizes       
$\frac{\sum_{i,j} A_{i,j} |y_i+y_j| }{\sum_i d_i |y_i|}$
  \end{itemize}
\end{itemize}
Using the construction of almost pairwise independent random variables
of Alon et al. \cite{AGHP92}, each sample space $\Omega_t$ has size $\tilde O(\log n)$,
and can be computed in $\tilde O(n)$ time. For each vector $y$, the
ratio can be computed in linear time. 
\end{proof}

By iterating the algorithm we derive our main result of this section.

\begin{theorem}
There is a nearly cubic time algorithm that in input a graph $G=(V,E)$ such that $max-cut-gain(G)\geq \epsilon$
finds a cut $(L,R)$ of $V$ of gain $\geq e^{-\Omega(1/\epsilon)}$ 
\end{theorem}

\section{Conclusions}

The motivating question for this work was to find a combinatorial interpretation
of the quantity $d-|\lambda_n|$ in a $d$-regular graph, akin to the interpretation
of $d-\lambda_2$ provided by the theory of edge expansion.

In establishing such an interpretation (in terms of the quantity that
we call ``bipartiteness ratio'' in Section \ref{sec:cheeger}) we proved
that a natural and easy-to-implement spectral algorithm performs non-trivially
well with respect to the Max Cut problem. 

The algorithm is very fast in practice \cite{OT08}; using a termination rule that
is slightly more relaxed than the one used in this paper (stopping when
$U+X > M/2$, instead of $U+X/2 > M/2$), the algorithm makes at most
one recursive call in all the experiments that we performed. It would
be interesting to give a proof that this is always the case.

A number of intersting open questions remain, such as:

\begin{enumerate}
\item What is the worst-case approximation ratio of our algorithm? We believe
that our bond $.531$ is not tight.

\item Is there a ``purely combinatorial'' algorithm
(namely, one not involving numerical matrix computations) 
for Max Cut achieving an approximation factor better than 1/2?

\item It should be possible to significantly improve our bounds for Max CutGain.

\end{enumerate}

\section*{Acknowledgements}

I would like to thank an anonymous commenter for asking the question of the
connection between spectral techniques and Max Cut,  and Sebastian Cioaba,
Satyen Kale, James Lee and Salil Vadhan
for providing  helpful comments and references to the related literature.

I am grateful to Moses Charikar for communicating the proof of Theorem \ref{th:largecut}, which
substantially improved my previous analysis, and for allowing me to present his
improved analysis in this paper.


\newpage

\appendix

\section{Appendix}

\subsection{Efficiency of the Arora-Kale Algorithm}
\label{sec:AK}

Arora and Kale \cite{AK07} describe an algorithm for the Goemans-Williamson SDP relaxation
of Max Cut which achieves an approximation ratio $1+o(1)$ and runs in time $\tilde O(D_{\max} \cdot |V|)$
given in input an unweighted multi-graph $G=(V,E)$  of maximum degree $D_{\max}$.\footnote{The
Arora-Kale result is more general, but this statement is sufficient for our purpose} In particular,
it is possible to find $(\alpha-o(1))$-approximate solutions to Max Cut in time  $\tilde O(D_{\max} \cdot |V|)$,
where $\alpha= .878\cdots$ is the approximation ratio of the Goemans-Williamson algorithm.

In this section we show that, using the Arora-Kale algorithm and a reduction from \cite{T01},
it is possible to approximate Max Cut within $\alpha-o(1)$ in time $\tilde O(|V|+|E|)$ regardless
of the degree distribution.\footnote{The running time can be reduced to $\tilde O(|V|)$ if
the representation of the graph is such that a random edge can be sampled in $\tilde O(1)$ time,
and the degree of a given vertex can be found in $\tilde O(1)$ time.}

Given the sparsification result discussed in Section \ref{sec:sparsify}, it is sufficient
to prove the following theorem, which is implicit in \cite{T01}.

\begin{theorem} There is a randomized algorithm $C$ and a deterministic algorithm $R$
with the following properties.

Given a graph $G=(V,E)$, algorithm $C$
 constructs in $\tilde O(|V|+|E|)$ time
a  graph $G'= (V',E')$ of maximum degree $\tilde O(1)$ with $|V'|= 2|E|$ vertices, such that
the following happens with high probability: (i) $maxcut (G') \geq maxcut (G) - o(1)$, and (ii)
given an arbitrary solution $S'\subseteq V'$
of cost $c$ in $G'$, algorithm $R$ constructs in  $\tilde O(|V|+|E|)$ time
a solution $S\subseteq V$ of cost $\geq c-o(1)$ for $G$.
\end{theorem}

\begin{proof}
We sketch how the argument in \cite{T01} applies to Max Cut.

Define the weighted graph $\hat G = (\hat V,\hat E)$ as follows. (This graph will only be used
in the analysis, and not explicitely constructed in the reduction.) For every vertex
$v\in V$ of degree $d_v$, $\hat V$ contains $d_v$ copies of $v$; for every edge $(u,v)$
in $E$, we have $d_u \cdot d_v$ edges $(\hat u,\hat v)$ in $E'$, one for every
copy $\hat u$ of $u$ and for every copy $\hat v$ of $v$, each such edge having
weight $1/(d_u \cdot d_v)$.

We claim that approximating Max Cut in $G$ is equivalent to approximating Max Cut in $\hat G$.
First, it should be clear that if $(S,V-S)$ is a cut in $G$ of cost $c$, then if we define
$\hat S\subseteq \hat V$ to be the set of all copies of vertices in $S$, then $(\hat S, \hat V- \hat S)$
is a cut of cost $c$ in $\hat G$. On the other hand, if $(\hat S, \hat V- \hat S)$ is a cut
of cost $c$, then consider the distribution over cuts in $G$ in which a vertex $v$ is
picked to be in $S$ with probability proportional to the fraction of copies of $v$
which are in $\hat S$; the expected fraction of cut edges in $G$ is exactly $c$, and using
the method of conditional expectations we can find a cut of cost at least $c$ in linear time.

The graph $G'$ is obtained by sampling with replacement $\tilde O(|\hat V|) = \tilde O(|V|+|E|)$ edges
from $\hat E$, using the distribution in which an edge is sampled with probability 
proportional to its weight. As discussed in Section \ref{sec:sparsify}, it follows from
Chernoff bounds that a solution of cost $c$ in $G'$ has cost $c \pm o(1)$ in $\hat G$.

It remains to discuss the complexity of sampling $G'$: to sample one edge, we first pick
a random edge $(u,v)$ of $G$, and then we pick at random one of the copies $\hat u$ of $u$
and one of the copies $\hat v$ of $v$; this distribution is equivalent to randomly sampling
one of the edges of $\hat G$ with probabiltiy proportional to its weight. After $O(|V|+|E|)$
time preprocessing, each edge of $G'$ can be sampled in constant time.\footnote{The point
of this discussion is that $\hat G$ may have $\Omega(|V|^2)$ edges even if $|E|= O(|V|)$,
for example if there are two vertices of degree $|V|-1$. This means that it is not possible to
explicitly construct $\hat G$ in $\tilde O(|V|+|E|)$ time, and so one must sample
edges from $\hat G$ without explicitly constructing $\hat G$.}

\end{proof}

\end{document}